\documentclass[a4paper,11pt]{article}
\pdfoutput=1 

\usepackage{jcappub} 

\usepackage[T1]{fontenc} 
\usepackage{placeins}
\usepackage[dvipsnames,svgnames,x11names,hyperref]{xcolor}
\usepackage{array}
\usepackage{graphicx}
\usepackage{booktabs}
\usepackage{slashed}
\RequirePackage{mathrsfs}
\usepackage{dsfont}
\usepackage{hyperref}
\hypersetup{
        unicode = true,
        pdftitle = Note , 
        colorlinks = true,
        linkcolor = Black!30!DodgerBlue4,
        citecolor = Black!30!DodgerBlue4,
        filecolor = Black!30!DodgerBlue4,
        urlcolor = Black!30!DodgerBlue4,
}

\usepackage{booktabs}
\usepackage{cancel}

\graphicspath{{./plots/}}

\usepackage{floatrow}
\newfloatcommand{capbtabbox}{table}[][\FBwidth]

\newcommand{\dd}{\mathrm{d}}

\renewcommand{\epsilon}{\varepsilon}

\renewcommand{\bar}{\overline}

\newcolumntype{L}{>{$}l<{$}}
\newcolumntype{C}{>{$}c<{$}}

\allowdisplaybreaks

\title{\boldmath Probing P- and CP-violation in  dark matter interactions}
\author[a]{Riccardo Catena}
\author[a]{Joakim Hagel}
\author[b]{and Carlos E.~Yaguna}

\affiliation[a]{Chalmers University of Technology, Department of Physics, SE-412 96 G\"oteborg, Sweden}
\affiliation[b]{Escuela de F\'isica, Universidad Pedag\'ogica y Tecnol\'ogica de Colombia, Avenida Central del Norte \# 39-115, Tunja, Colombia}

\emailAdd{catena@chalmers.se}
\emailAdd{hagel@student.chalmers.se}
\emailAdd{carlos.yaguna@uptc.edu.co}

\abstract{
Discrete symmetries played a central role in elucidating the structure of the weak interactions, and they will probably be equally crucial regarding the interactions of the dark matter (DM) particle -- whose nature remains elusive. In this work we show that signals in future direct detection experiments can be used to test, in a model-independent way, for P- and CP-violation in DM-nucleus interactions.~The analysis is performed within the most general effective theory for non-relativistic spin-0 DM-nucleus interactions mediated by the exchange of a heavy particle.~Assuming an idealised xenon detector, we calculate the expected number of DM signal events  required to reject P and CP invariant DM-nucleus interactions. For a DM mass of 30~GeV (or higher), this number lies between about 10 and 300 DM signal events, depending on how  P and CP invariance are modeled. Future direct detection experiments, therefore, have the potential to reveal P- and CP-violation in DM interactions, making a decisive step toward the identification of the DM particle.
} 

\begin{document} 
\maketitle
\flushbottom

\section{Introduction}
One of the priorities in astroparticle physics is detecting the particles forming the cosmological dark matter (DM) component and interpreting this discovery in terms of particle physics models~\cite{appec}.~Direct detection experiments~\cite{Drukier:1983gj} are expected to play a major role in this context~\cite{Undagoitia:2015gya}.~They search for nuclear recoils~\cite{Goodman:1984dc} (or electronic transitions~\cite{Essig:2011nj}) induced by the non-relativistic scattering of DM particles from the Milky Way in low-background detectors located deep underground.~Next generation direct detection experiments that will operate double-phase xenon time projection chambers, such as XENONnT~\cite{Aprile:2015uzo} and LZ~\cite{Akerib:2018lyp}, are expected to probe standard spin-independent DM-nucleon scattering cross sections of approximately 10$^{-48}$~cm$^2$ for DM particle masses around 50 GeV, and to reach exposures of about 20 ton$\times$year~\cite{Aprile:2015uzo,Akerib:2015cja}.~Based on this expected performance, if DM is made of weakly interacting massive particles (WIMPs), a DM particle discovery at direct detection experiments is within reach.

In this work we are interested in what can be learned about the DM particle properties from a positive signal in future direct detection experiments such as XENONnT~\cite{Aprile:2015uzo}, LZ~\cite{Akerib:2018lyp}, or DARWIN~\cite{Aalbers:2016jon}.~While the standard approach to the analysis of results from DM direct detection experiments focuses on constraining the DM particle mass and cross section, there has recently been a mounting interest in exploiting the direct detection technique to address more ambitious questions.~These include assessing whether the DM is its own antiparticle or not~\cite{Queiroz:2016sxf,Kavanagh:2017hcl,Catena:2020tbv}, extracting the DM particle spin~\cite{Catena:2017wzu,Baum:2017kfa}, and simultaneously measuring the DM-nucleon scattering cross section and the local DM density~\cite{Kavanagh:2020cvn}.

The main purpose of this work is to determine if a positive signal at direct detection experiments can shed some light on whether the interactions between DM and nuclei are parity (P) and charge conjugation-parity (CP) preserving or not -- a result with major implications on the fundamental nature of the DM.~While this question is too general to be addressed in the case of DM candidates of arbitrary spin, it can conveniently be formulated in terms of a simple hypothesis test when focusing on scalar (spin-0) DM.~Specifically, if DM has spin 0, there are only four possible ways, or interaction operators, to couple DM to the nucleons bound in target nuclei.~Two of them  preserve both P and CP, one violates P but preserves CP and the last one violates P and CP. These operators predict signals with  distinct energy spectra so it is feasible to differentiate them and in this way  establish P- and CP-violation. ~Starting from this consideration, we define a ``null hypothesis'' corresponding to P and CP preserving scalar DM-nucleon interactions and two alternative hypotheses according which DM has spin 0 and interactions violating P, or P and CP, respectively.~We then use the likelihood ratio test statistic to compute the number of signal events that a xenon experiment has to record in order to reject the null hypothesis (P and CP preserving interactions) in favour of one of the alternative hypotheses at a given statistical significance.~We find that the number of signal events required to reject the null hypothesis with a statistical significance corresponding to three standard deviations varies from about 10 to 300 depending on the specific hypotheses we compare. This means that future direct detection experiments may indeed reveal P- or CP-violation in DM interactions, a discovery of crucial significance for  the identification of the DM particle.

This paper is organised as follows.~In Sec.~\ref{sec:theory} we review the theory of DM scattering by nuclei.~Specifically, we focus on the transformation properties under P and CP of the most general non-relativistic amplitude for DM-nucleon scattering.~Here, we also review the formalism of DM direct detection.~In Sec.~\ref{sec:stat}, we introduce the statistical methods we use to test the null hypothesis (P and CP preserving interactions) against two alternative hypothesis corresponding to P, or P- and CP-violation, respectively.~We present our results in terms of number of signal events required to reject the null hypothesis in Sec.~\ref{sec:results}, and finally conclude in Sec.~\ref{sec:conclusions}.~In Appendix~\ref{sec:T}, we provide a detailed derivation for some of the results reported in Sec.~\ref{sec:theory}.

\section{P and CP invariance in spin-0 dark matter-nucleus scattering}
\label{sec:theory}
In this section we review the theory of DM scattering by atomic nuclei and its application to DM direct detection.~Focusing on the most general non-relativistic effective theory for spin-0 DM-nucleon interactions mediated by a heavy particle~\cite{Dobrescu:2006au,Fan:2010gt,Fitzpatrick:2012ix}, in Sec.~\ref{sec:S} we classify the transformation properties under P and CP of the S-matrix and cross section for DM-nucleus scattering while reviewing the formalism and basic assumptions of the DM direct detection technique in Sec.~\ref{sec:dd}.~We provide a derivation for some of the expressions reported here in Appendix~\ref{sec:T}.

\subsection{Effective theory expansion of the S-matrix and cross section}
\label{sec:S}

In Appendix~\ref{sec:T}, we derive the following expression for the T-matrix element, $T_{fi}$, for DM-nucleus scattering,
\begin{align}
T_{fi} = -\sum_{j=1}^{A} \left( \prod_{i=1}^{A} \int \frac{{\rm d}^3 \mathbf{k}_i}{(2 \pi)^3} \right) \frac{\psi^*_f(\mathbf{k}_1^j,
\dots,\mathbf{k}^j_A
) \psi_{in}(\mathbf{k}_1,\dots, \mathbf{k}_A)}{\sqrt{16 E_{\mathbf{p}} E_{\mathbf{k}_j} E_{\mathbf{p}'} E_{\mathbf{k}'_j}}}\,M_{\chi N_j}\,,
\label{eq:Tfinal_sec2}
\end{align}
where $\mathbf{k}_m^j=\mathbf{k}_m+\mathbf{q}$ for $m=j$ and $\mathbf{k}_m^j=\mathbf{k}_m$ otherwise.~Here, $\mathbf{q}=\mathbf{p}-\mathbf{p}'$ is the momentum transfer, $\mathbf{p}$ ($\mathbf{p}'$) is the initial (final) DM particle momentum, while $\mathbf{k}_1, \dots, \mathbf{k}_A$ are nucleon momenta.~Furthermore, $E_{\mathbf{p}}$ ($E_{\mathbf{p}'}$) and $E_{\mathbf{k}}$ ($E_{\mathbf{k}'}$) are the initial (final) DM and nucleon energies, while we denote by $\psi_{in}$ and  $\psi_f$ the incoming and outgoing nuclear wave functions, respectively.~In order to simplify the notation, in Eq.~(\ref{eq:Tfinal_sec2}), we omit the nucleon spin indices (see Appendix~{\ref{sec:T}} for further details).~Assuming that nuclear ground states are eigenstates of P and CP so that $\psi_f^* \psi_{in}$ is even under these transformations, Eq.~(\ref{eq:Tfinal_sec2}) shows that the S-matrix element $S_{fi}$ transforms under P and CP in a way that is entirely determined by the amplitude for DM scattering by a free nucleon, $M_{\chi N_j}$.~In the non-relativistic limit, $M_{\chi N_j}$ can be written as a function of $\mathbf{q}$ and $\mathbf{v}_j^\perp=(\mathbf{p}+\mathbf{p}')/(2 m_\chi) - (\mathbf{k}_j+\mathbf{k}_j')/(2 m_N)$ solely, as out of the four three-dimensional momenta, $\mathbf{p}$, $\mathbf{p}'$, $\mathbf{k}_j$ and $\mathbf{k}_j'$ only two are independent~\cite{Dobrescu:2006au,Fan:2010gt,Fitzpatrick:2012ix} due to momentum conservation and the requirement that $M_{\chi N_j}$ is invariant under Galilean transformations, i.e.~constant shifts of particle velocities.~At first order in $|\mathbf{q}|/m_{N}$ and at first order in $\mathbf{v}^\perp_j$, the most general form for the scattering amplitude $M_{\chi N_j}$ in the case of spin-0 DM is~\cite{Fitzpatrick:2012ix}~\footnote{In the case of spin-1/2 and spin-1 DM, the full non-relativistic expansion of $M_{\chi N_j}$ also involves terms quadratic in $|\mathbf{q}|/m_{N}$~\cite{Catena:2019hzw,DelNobile:2018dfg}.},
\begin{align}
M_{\chi N_j} &= \sum_{\ell =1,3,7,10} c_{\ell}^N \xi^{\dagger s'}_j\mathcal{O}_{\ell}\xi^{s}_j \,,
\end{align}
where $m_N$ is the nucleon mass and the DM-nucleon interaction operators $\mathcal{O}_{1}$,  $\mathcal{O}_{3}$, $\mathcal{O}_{7}$ and $\mathcal{O}_{10}$ are defined via~\cite{Fitzpatrick:2012ix},
\begin{align}
M_{\chi N_j} &\equiv c_{1}^N\xi^{\dagger s'}_j\mathds{1}_{N_j}\xi^s_j + c_{3}^N \,i \xi^{\dagger s'}_j\mathbf{S}_{N_j}\xi^s_j\cdot \left( \frac{\mathbf{q}}{m_N} \times \mathbf{v}^\perp_j \right)+ c_{7}^N \,\xi^{\dagger s'}_j\mathbf{S}_{N_j}\xi^s_j\cdot  \mathbf{v}^\perp_j \nonumber\\
&+ c_{10}^N \,i\xi^{\dagger s'}_j\mathbf{S}_{N_j}\xi^s_j\cdot \left( \frac{\mathbf{q}}{m_N}\right) \,.
\label{eq:M}
\end{align}
Here, $\xi^{\dagger s'}_j$ and $\xi^s_j$ are two-component spinors arising from the non-relativistic expansion of nucleon fields, $\mathbf{S}_{N_j}$ is the nucleon spin and $\mathds{1}_{N_j}$ the $2\times2$ identity matrix.~For the interaction operators $\mathcal{O}_{\ell}$, $\ell=1,3,7,10$, we follow the notation introduced in~\cite{Fitzpatrick:2012ix}.~Contrary to~\cite{Fitzpatrick:2012ix}, within our conventions the coupling constants $c_{1}^N$, $c_{3}^N$, $c_{7}^N$ and $c_{10}^N$ in Eq.~(\ref{eq:M}) are dimensionless.~In the numerical applications, we assume that they are positive and the same for protons and neutrons.

The transformation properties of $M_{\chi N_j}$ (and thus of $S_{fi}$) under P and CP are therefore determined by those of the building blocks $\mathbf{q}$, $\mathbf{S}_{N_j}$ and $\mathbf{v}^\perp_j$.~The spin operator, $\mathbf{S}_{N_j}$, is even under P and odd under CP, the transverse relative velocity $\mathbf{v}^\perp_j$\footnote{In the case of elastic DM-nucleon scattering, the velocity $\mathbf{v}^\perp_j$ is transverse with respect to the momentum transfer $\mathbf{q}$, i.e.~$\mathbf{v}^\perp_j\cdot \mathbf{q}=0$} is odd under both P and CP, while the momentum transfer $\mathbf{q}$ is odd under P and even under CP.~Based on these transformation properties, we conclude that $\mathcal{O}_1 = \mathds{1}_{N_j}$ and $\mathcal{O}_3 = \,i \mathbf{S}_{N_j}\cdot [(\mathbf{q}/m_N) \times \mathbf{v}^\perp_j ]$ preserve P and CP,  $\mathcal{O}_7=\mathbf{S}_{N_j}\cdot  \mathbf{v}^\perp_j$ violates P but preserves CP while $\mathcal{O}_{10}=\,i\mathbf{S}_{N_j}\cdot (\mathbf{q}/m_N)$ violates both.

The results reviewed in this section allow us to express the differential cross section for DM scattering by an atomic nucleus in terms of the T-matrix element in Eq.~(\ref{eq:Tfinal_sec2}).~It reads as follows
\begin{align}
\frac{{\rm d} \sigma_T}{{\rm d} E_R} = \frac{m_T}{2\pi v^2} \, \frac{1}{2J_T+1} \sum_{\rm spins}|T_{fi}|^2\,,
\label{eq:sigma}
\end{align}
where $v$ is the DM-nucleus relative velocity, $E_R=|\mathbf{q}|^2/(2 m_T)$, $m_T$ and $J_T$ are the nuclear recoil energy, mass and spin, respectively, and the sum runs over initial and final nuclear spin configurations.

\subsection{Expected rate of nuclear recoils in underground detectors}
\label{sec:dd}
The expected differential rate of nuclear recoils per unit detector mass in a direct detection experiment,
\begin{align}
\frac{{\rm d}\mathscr{R}}{{\rm d}E_R} = E \frac{\rho_\chi}{m_\chi} \sum_{T} \frac{\xi_T}{m_T} \int_{|\mathbf{v}|>v_{\rm min}} {\rm d}^3 v \,|\mathbf{v}| f(\mathbf{v}+\mathbf{v}_{\oplus})\,\frac{{\rm d} \sigma_T}{{\rm d} E_R}
\label{eq:R}
\end{align}
depends on the local DM density $\rho_\chi$, the DM velocity distribution in the rest frame of our galaxy boosted to the detector rest frame, $f(\mathbf{v}+\mathbf{v}_{\oplus})$, where $\mathbf{v}_{\oplus}$ is the Earth's velocity with respect to the galactic center, and on the differential cross section for DM-nucleus scattering in Eq.~(\ref{eq:sigma}).~In Eq.~(\ref{eq:R}), the sum is performed over the most abundant nuclear targets forming the detector material, $\xi_T$ is the mass fraction of the target $T$ and $v_{\rm min}=|\mathbf{q}|/(2\mu_T)$, where $\mu_T$ is the reduced DM-nucleus mass, is the minimum velocity required to transfer a momentum $|\mathbf{q}|$ or,  equivalently, an energy $E_R$ in the DM-nucleus scattering process.~Finally, $m_\chi$ is the DM mass and $E$ is the detector efficiency, i.e.~the expected fraction of signal events passing all experimental cuts.

In the numerical applications, we set $\rho_\chi=0.3$~GeV~cm$^{-3}$ and assume for $f$ a Maxwell-Boltzmann distribution truncated at the escape velocity $v_{\rm esc}=550$~km~s$^{-1}$ with most probable speed equal 220~km~s$^{-1}$, as in the so-called Standard Halo Model~\cite{Freese:2012xd}.~As far as the detector material is concerned, we focus on xenon and evaluate the differential cross section in Eq.~(\ref{eq:sigma}) by using the \texttt{DMFormFactor} code~\cite{Anand:2013yka}, which can account for the most abundant xenon isotopes.~We assume an energy independent detector efficiency and set $E$ to the constant value 0.7.

\begin{table}
\renewcommand{\arraystretch}{1.8}
\centering
\begin{tabular}{lllll}
\hline
Symbol & Properties & Operators & Hierarchy & Label \\
\hline
\hline
$\mathcal{H}_{0}^{(1)}$ & P and CP preserving & $c_1^N \mathcal{O}_1 + c_3^N \mathcal{O}_3$ & $\frac{(c_3^N)^2\mathcal{N}^{33}}{(c_1^N)^2 \mathcal{N}^{11}}=10^{-3}$ & ``$\mathcal{O}_1$ Tyranny''  \\
$\mathcal{H}_{0}^{(2)}$ & P and CP preserving & $c_1^N \mathcal{O}_1 + c_3^N \mathcal{O}_3$ & $\frac{(c_1^N)^2\mathcal{N}^{11}}{(c_3^N)^2 \mathcal{N}^{33}}=10^{-3}$ & ``$\mathcal{O}_3$ Tyranny''  \\
$\mathcal{H}_{0}^{(3)}$ & P and CP preserving & $c_1^N \mathcal{O}_1 + c_3^N \mathcal{O}_3$ & $\frac{(c_3^N)^2\mathcal{N}^{33}}{(c_1^N)^2 \mathcal{N}^{11}}=1$ & ``Democracy'' \\
$\mathcal{H}_{A1}$       & P violating  & $c_7^N \mathcal{O}_7$ & &\\
$\mathcal{H}_{A2}$ 	     & P and CP violating & $c_{10}^N \mathcal{O}_{10}$ & & \\
\hline
\end{tabular}
\caption{List of hypotheses, including the corresponding symbols, properties, operators, hierarchies and labels.}
\label{tab:h}
\end{table}

\section{Statistical methods for hypothesis testing}
\label{sec:stat}
In this section we describe the statistical methods that we use to compare a P and CP preserving hypothesis against P, or P and CP, violating models.~In Sec~\ref{sec:h}, we formulate the hypotheses that we are interested in comparing.~In Sec.~\ref{sec:l} and Sec.~\ref{sec:mc}, respectively, we introduce the test statistic and Monte Carlo simulations that we use to perform such a comparison.

\subsection{P and CP preserving/violating hypotheses}
\label{sec:h}
In our analysis we compare a ``null hypothesis'' where DM interacts with atomic nuclei via P and CP preserving interactions, $\mathcal{H}_0$, with alternative hypotheses where P, or P and CP are violated.~In all cases considered in this work, the DM is assumed to consist of spin-0 particles.~Even within the assumption of spin-0 DM, there is not a unique P and CP preserving model for DM-nucleus interactions, as both the operator $\mathcal{O}_1$ and the operator $\mathcal{O}_3$ are even under P and CP.~This implies that when P and CP are conserved in spin-0 DM models, the scattering amplitude $M_{\chi N_j}$ is in general a linear combination of the $\mathcal{O}_1$ and $\mathcal{O}_3$ interaction operators, and the expected rate of nuclear recoil events in the energy interval, $\Delta E_R$, can be written as
\begin{align}
\mathcal{N} = \mathcal{E}\int_{\Delta E_R} {\rm d}E_R\,\frac{{\rm d} \mathscr{R}}{{\rm d} E_R}  = (c_1^N)^2 \mathcal{N}^{11} + (c_3^N)^2\mathcal{N}^{33} +(2 c_1^N c_3^N)\gamma \mathcal{N}^{13}\,,
\label{eq:N}
\end{align}
where $\mathcal{E}$ is the exposure.~In Eq.~(\ref{eq:N}), we introduced $\mathcal{N}^{11}=\mathcal{N}|_{c_1^N=1, c_3^N=0}$, $\mathcal{N}^{33}=\mathcal{N}|_{c_1^N=0, c_3^N=1}$ and $\mathcal{N}^{13}=(\mathcal{N}|_{c_1^N=1, c_3^N=1}-\mathcal{N}^{11}-\mathcal{N}^{33})/2$ to make the dependence of $\mathcal{N}$ on $c_1^N$ and $c_3^N$ explicit.~Since the relative contribution to the expected rate of nuclear recoils from $\mathcal{O}_1$ and  $\mathcal{O}_3$ remains unspecified within the hypothesis of C and CP preserving interactions, we consider three different scenarios separately.~In a first scenario, $\mathcal{H}_0^{(1)}$, the nuclear recoil rate is dominated by the $\mathcal{O}_1$ operator, i.e.~$(c_3^N)^2\mathcal{N}^{33}/(c_1^N)^2 \mathcal{N}^{11}=10^{-3}$.~We refer to this scenario as ``$\mathcal{O}_1$ Tyranny''.~The second scenario, $\mathcal{H}_0^{(2)}$, corresponds to models where $\mathcal{O}_3$ gives the largest contribution to the expected rate of nuclear recoils, i.e.~$(c_1^N)^2\mathcal{N}^{11}/(c_3^N)^2 \mathcal{N}^{33}=10^{-3}$.~We refer to this scenario as ``$\mathcal{O}_3$ Tyranny''.~Finally, in the third scenario, $\mathcal{H}_0^{(3)}$', the $\mathcal{O}_1$ and $\mathcal{O}_3$ contribution to the recoil rate is such that $(c_1^N)^2\mathcal{N}^{11}/(c_3^N)^2 \mathcal{N}^{33}=1$.~We refer to this scenario as ``Democracy''.~The three realisations of $\mathcal{H}_0$, $\mathcal{H}_0^{(i)}$, $i=1,2,3$, are summarised in Tab.~\ref{tab:h}.~While considering three versions of $\mathcal{H}_0$ allows us to draw rather general conclusions, it is important to notice that in concrete models for spin-0 DM, $\mathcal{O}_1$ is expected to generically dominate over  $\mathcal{O}_3$~\cite{Baum:2017kfa}.

We compare two alternative hypotheses against each of the three versions of $\mathcal{H}_0$.~In the first alternative hypothesis, DM interacts with atomic nuclei via P violating (but CP preserving) interactions and $M_{\chi N_j}$ is proportional to the $\mathcal{O}_7$ operator.~The second alternative hypothesis corresponds to models where DM interacts with atomic nuclei via P and CP violating interactions and $M_{\chi N_j}$ is proportional to $\mathcal{O}_{10}$.~We denote the former by $\mathcal{H}_{A1}$ and the latter by $\mathcal{H}_{A2}.$~Tab.~\ref{tab:h} summarises the hypotheses that we compare and the corresponding properties.

\subsection{Log-likelihood ratio}
\label{sec:l}
We compare the three versions of $\mathcal{H}_0$ with $\mathcal{H}_{A1}$ and $\mathcal{H}_{A2}$ using the log-likelihood ratio test statistic~\cite{Cowan:2010js},
\begin{align}
t = -2 \ln \frac{\max_{\boldsymbol{\Theta}\in\Omega_{\rm null}}\mathscr{L}(\mathscr{D}|\boldsymbol{\Theta})}{\max_{\boldsymbol{\Theta}\in\Omega_{\rm alter}}\mathscr{L}(\mathscr{D}|\boldsymbol{\Theta})} \,.
\label{eq:t}
\end{align}
In Eq.~(\ref{eq:t}), $\boldsymbol{\Theta}=\{\theta_1\equiv c_1^N, \theta_2\equiv c_3^N, \theta_3\equiv c_7^N,  \theta_4\equiv c_{10}^N\}$ and $\Omega_{\rm null}=\{\boldsymbol{\Theta}~:~\theta_1\ge 0, \theta_2\ge 0, \theta_3=0, \theta_4=0~|~\Gamma=0\}$, where 
$\Gamma=\theta_2^2\mathcal{N}^{33}/\theta_1^2 \mathcal{N}^{11}-10^{-3}$ for $\mathcal{H}_0=\mathcal{H}^{(1)}_0$,  $\Gamma=\theta_1^2\mathcal{N}^{11}/\theta_2^2 \mathcal{N}^{33}-10^{-3}$ for $\mathcal{H}_0=\mathcal{H}^{(2)}_0$ and $\Gamma=\theta_2^2\mathcal{N}^{33}/\theta_1^2 \mathcal{N}^{11}-1$ for $\mathcal{H}_0=\mathcal{H}^{(3)}_0$.~Furthermore, $\Omega_{\rm alter}=\{\boldsymbol{\Theta}~:~\theta_1=0,  \theta_2=0, \theta_3\ge0, \theta_4=0 \}$ in the case of $\mathcal{H}_{A1}$ and  $\Omega_{\rm alter}=\{\boldsymbol{\Theta}~:~\theta_1=0,  \theta_2=0, \theta_3=0, \theta_4\ge0 \}$ in the case of $\mathcal{H}_{A2}$.~Finally, we denote by $\mathscr{D}=\{\mathscr{N}_1, \dots \mathscr{N}_n\}$ the dataset of hypothetically observed nuclear recoil energies in $n$ independent energy bins covering the signal region, $\Delta E_R$.~We generate the dataset $\mathscr{D}$ via Monte Carlo simulations, as explained in Sec.~\ref{sec:mc}.~Our reference value for $m_\chi$ in these simulations is $30$~GeV.~We comment on this assumption in Sec.~\ref{sec:results}.

For each energy bin $\Delta E_R^{(i)}$, $i=1,\dots,n$, we assume a Poisson likelihood, so that the total likelihood is
\begin{equation}
\mathscr{L}(\mathscr{D}|\boldsymbol{\Theta}) = \prod_{i=1}^{n} \frac{\left(\mathscr{B}_{i}+\mathscr{S}_{i}(\boldsymbol{\Theta})\right)^{\mathscr{N}_{i}}}{\mathscr{N}_{i}!} e^{-\left(\mathscr{B}_{i}+\mathscr{S}_{i}(\boldsymbol{\Theta})\right)} \,,
\end{equation}
where 
\begin{equation}
\mathscr{S}_{i}(\boldsymbol{\Theta}) = \mathcal{E} \int_{\Delta E_R^{(i)}} {\rm d} E_R \, \frac{\dd\mathscr{R}}{\dd E_R} \,,
\label{eq:S_i}
\end{equation}
and, as anticipated, $ \mathcal{E}$ is the experimental exposure.~Here, $\mathscr{B}_{i}$ is the expected number of background events in the $i$-th energy bin.~For the purposes of this study, we set $\mathscr{B}_{i}=0$, since we expect $\mathscr{S}_{i}$ to be significantly larger than $\mathscr{B}_{i}$ when one of the realisations of $\mathcal{H}_0$ can be rejected in favour of $\mathcal{H}_{
A1}$ or $\mathcal{H}_{A2}$.~This at least in a sufficiently large number of energy bins.~In all numerical applications, we assume the signal region $[5, 50]$~keV, $n=20$, and $\Delta E_R^{(i)}=45/20$~keV.

\subsection{Monte Carlo simulations}
\label{sec:mc}
For each of the hypotheses $\mathcal{H}_0^{(1)}$, $\mathcal{H}_0^{(2)}$, $\mathcal{H}_0^{(3)}$, $\mathcal{H}_{A1}$ and $\mathcal{H}_{A2}$, we sample the dataset $\mathscr{D}$ from $n$ Poissonians of mean given by Eq.~(\ref{eq:S_i}) with $\boldsymbol{\Theta}=\boldsymbol{\Theta}'$, where, depending on the underlying hypothesis, 
\begin{align}
\boldsymbol{\Theta}'&
=\left\{\theta_1=\alpha,  \theta_2=\beta, \theta_3=0, \theta_4=0~|~\Gamma(\alpha,\beta)=0 \right\} \,, \qquad \text{for} \,\,\mathcal{H}_0 \,, \label{eq:benchmark H0}\, \\
\boldsymbol{\Theta}'&
=\left\{\theta_1=0,  \theta_2=0, \theta_3=\gamma, \theta_4=0 \right\} \,, \qquad\qquad\qquad\qquad\, \text{for} \,\,\mathcal{H}_{A1} \,,
\label{eq:benchmark HA1} \\
\boldsymbol{\Theta}'&
=\left\{\theta_1=0,  \theta_2=0, \theta_3=0, \theta_4=\delta \right\} \,, \qquad\qquad\qquad\qquad\, \text{for} \,\,\mathcal{H}_{A2} \,.
\label{eq:benchmark HA2}
\end{align}
For a given hypothesis and experimental exposure, $\mathcal{E}$, we can vary $\alpha$, $\beta$, $\gamma$ or $\delta$ to obtain the desired number of expected signal events, $\mathscr{S}_{\rm tot}=\sum_{i=1}^n \mathscr{S}_i(\boldsymbol{\Theta}')$.~Alternatively, one can fix $\alpha$, $\beta$, $\gamma$ or $\delta$ to a reference value and then vary the experimental exposure in order to obtain the desired value for $\mathscr{S}_{\rm tot}$.~Indeed, for a given hypothesis $\mathscr{S}_{\rm tot}$ depends on $\mathcal{E}$ times a squared coupling and a change in the former can always be compensated by a change in the latter.~Finally, notice also that because of the constraint $\Gamma(\alpha,\beta)=0$, $\alpha$ and $\beta$ are not independent.

\begin{figure}
\centering
\includegraphics[width=0.77\textwidth]{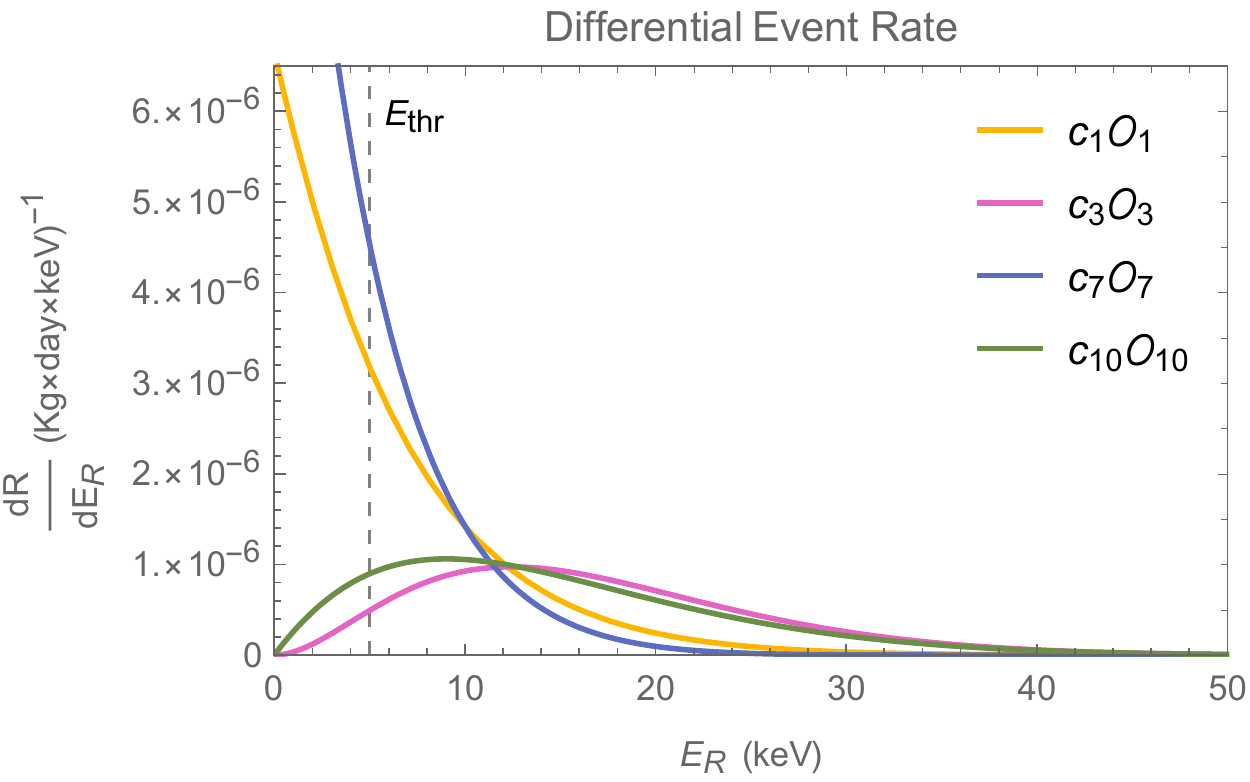}
\caption{Nuclear recoil energy spectra predicted for the $\mathcal{O}_1$, $\mathcal{O}_3$, $\mathcal{O}_7$ and $\mathcal{O}_{10}$ operators in a xenon detector.~We set the DM particle mass to 30 GeV, and the coupling constant $c_1^N$ to a value corresponding to a DM-nucleon scattering cross section of 4.1$\times 10^{-47}$~cm$^2$.~We set the coupling constants for the other operators to values producing the same number of signal events in the [5,50]~keV energy range as $\mathcal{O}_1$ with parameters set as above.~The dashed, vertical line indicates the assumed energy threshold.}
\label{fig:spectra}
\end{figure}

By repeatedly sampling $\mathscr{D}$ under (one of the realisations of) $\mathcal{H}_0$, we obtain the probability density function of $t$ under (that realisation of) $\mathcal{H}_0$, which we denote by~$f(t|\mathcal{H}_0)$.~Similarly, by repeatedly simulating $\mathscr{D}$ under $\mathcal{H}_{A1}$ or $\mathcal{H}_{A2}$, we obtain the probability density function of $t$ under $\mathcal{H}_{A1}$ or, analogously, under $\mathcal{H}_{A2}$.~We denote these by $f(t|\mathcal{H}_{A1})$ and $f(t|\mathcal{H}_{A2})$, respectively.~The significance for rejecting P and CP preserving DM-nucleon interactions in spin-0 DM models in favour of P ($\mathcal{H}_{A1}$), or P and CP ($\mathcal{H}_{A2}$), violating interactions, $\mathcal{Z}$, is then
\begin{equation}
\mathcal{Z} = \Phi^{-1}(1-p)\,,
\label{eq:Z}
\end{equation}
where $\Phi$ is the cumulative distribution function of a Gaussian probability density of mean 0 and variance 1,
\begin{equation}
p = \int^{\infty}_{t_{\rm med}} {\rm d}t\, f(t|\mathcal{H}_{A1/A2})\,,
\label{eq:pvalue}
\end{equation}
is the $p$-value for rejecting (one of the realisations of) $\mathcal{H}_{0}$ in favour of $\mathcal{H}_{A1}$ or $\mathcal{H}_{A2}$, and  $t_{\rm med}$ is the median of $f(t|\mathscr{H}_{A1})$ in the former case, and the median of $f(t|\mathscr{H}_{A2})$ in the latter.~In order to obtain the results presented in Sec.~\ref{sec:results}, we compute $f(t|\mathcal{H}_{0})$ from about $2\times10^{4}$ Monte Carlo simulations of $t$ under $\mathcal{H}_0$, while we find $t_{\rm med}$ by sampling about $2\times10^4$ values for $t$ under $\mathcal{H}_{A1}$ or  $\mathcal{H}_{A2}$ and then computing the median of this sample.~Notice that since $\mathcal{H}_{0}$, $\mathcal{H}_{A1}$ and $\mathcal{H}_{A2}$ are not nested hypotheses (i.e.~they do not coincide when setting to zero a subset of parameters), we cannot rely on asymptotic formulae for the probability density function of the test statistic $t$, and Monte Carlo simulations are required to obtain both $f(t|\mathcal{H}_{0})$ and $f(t|\mathcal{H}_{A1/A2})$.

\begin{figure}
\centering
\includegraphics[width=0.77\textwidth]{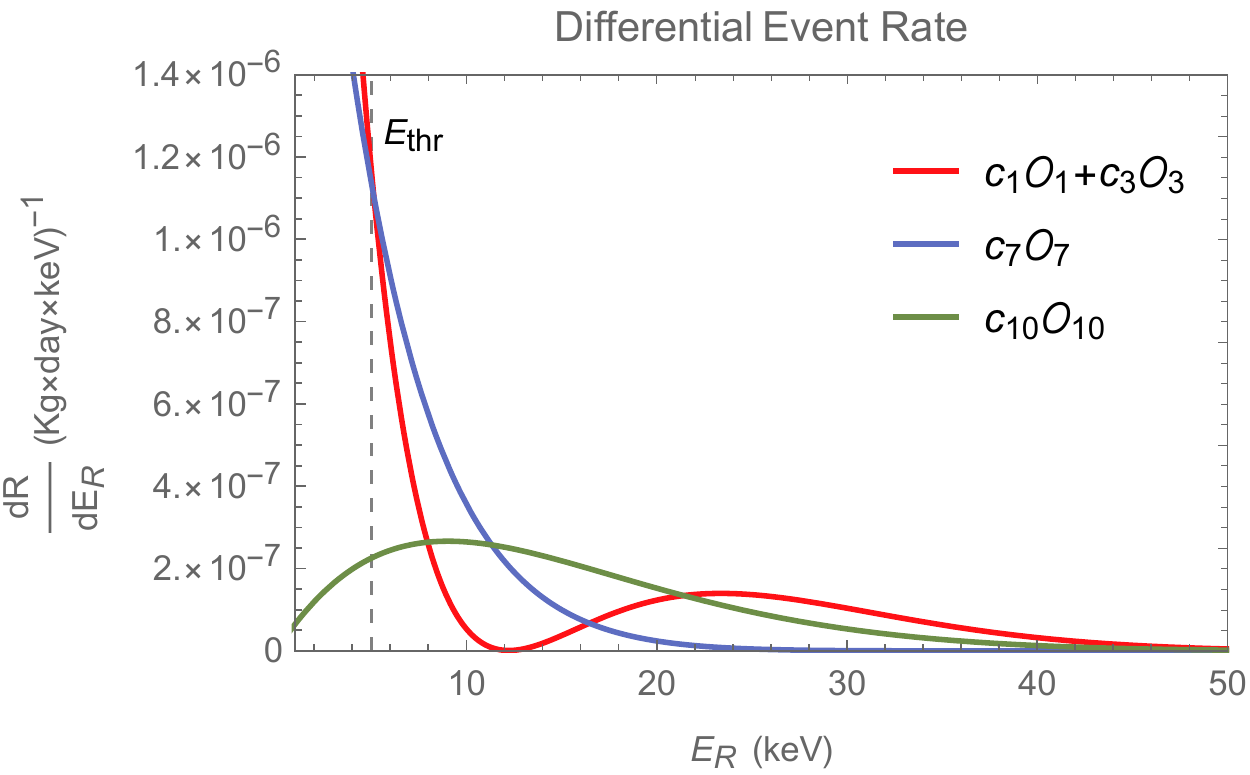}
\caption{Same as Fig.~\ref{fig:spectra}, but now comparing the $\mathcal{H}_{A1} (\mathcal{O}_7)$ and $\mathcal{H}_{A2} (\mathcal{O}_{10})$ hypotheses with $\mathcal{H}_0^{(3)}$ (``Democracy''), see Tab.~\ref{tab:h}.}
\label{fig:spectra2}
\end{figure}

\section{Results}
\label{sec:results}

\subsection{General considerations}
In this section, we compute the number of DM signal events required to reject the three realisations of $\mathcal{H}_0$ that we consider ($\mathcal{O}_1$ Tyranny, $\mathcal{O}_3$ Tyranny, Democracy) in favour of $\mathcal{H}_{A1}$ or $\mathcal{H}_{A2}$, separately.~The possibility of rejecting P and CP invariant spin-0 DM-nucleus interactions with direct detection experiments relies on the difference between the predicted nuclear recoil energy spectra under (the three realisations of) $\mathcal{H}_0$, $\mathcal{H}_{A1}$ and $\mathcal{H}_{A2}$, respectively.~In order to gain intuition for when rejecting $\mathcal{H}_0$ is expected to be simple and when it is expected to be  challenging, in Fig.~\ref{fig:spectra} we report the nuclear recoil energy spectra associated with the four interaction operators $\mathcal{O}_1$, $\mathcal{O}_3$, $\mathcal{O}_7$ and $\mathcal{O}_{10}$, respectively, whereas Fig.~\ref{fig:spectra2} shows the predicted spectra under the democratic realisation of $\mathcal{H}_0$, $\mathcal{H}_{A1}$ and $\mathcal{H}_{A2}$ with coupling constants set to produce the same number of signal events in the signal region $\Delta E_R=[5, 50]$~keV.~While the $\mathcal{O}_1$ and $\mathcal{O}_7$ operators are characterised by an exponentially decaying nuclear recoil energy spectrum, the spectra associated with $\mathcal{O}_3$ and $\mathcal{O}_{10}$ exhibit a peak at a finite value of $E_R$.~The exact location of the peak depends on the DM particle mass and moves towards larger values of $E_R$ when increasing $m_\chi$.~Based on these considerations, we expect that the $\mathcal{O}_1$ Tyranny should be difficult to reject in favour of P violating spin-0 DM-nucleus interactions.~The same applies to the $\mathcal{O}_3$ Tyranny and P and CP violating spin-0 DM-nucleus interactions.~In contrast, $\mathcal{O}_3$ Tyranny ($\mathcal{H}_0^{(2)}$) and $\mathcal{H}_{A1}$ (corresponding to P-violations) should be simple to discriminate.

\subsection{Rejecting P and CP invariance quantitatively}
\begin{figure}
\centering
\includegraphics[width=0.49\textwidth]{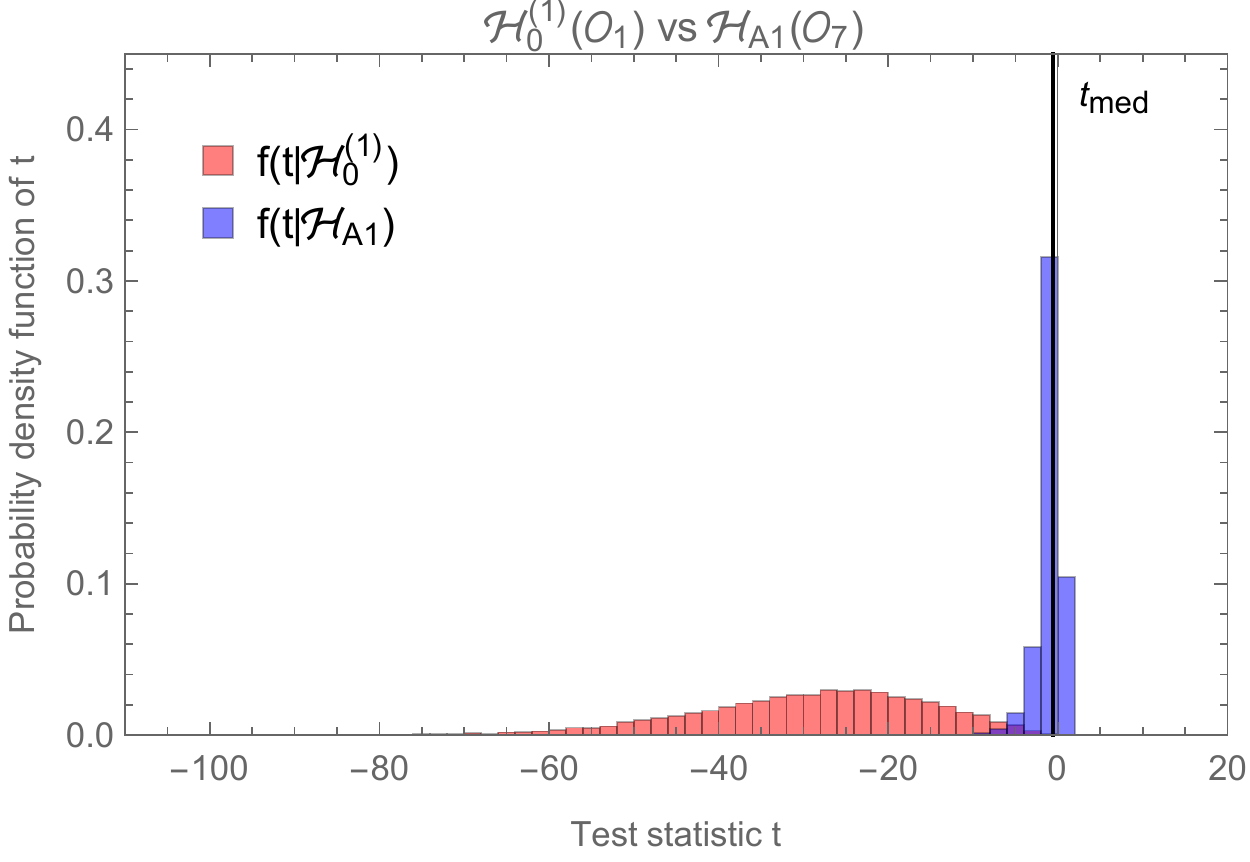}
\includegraphics[width=0.49\textwidth]{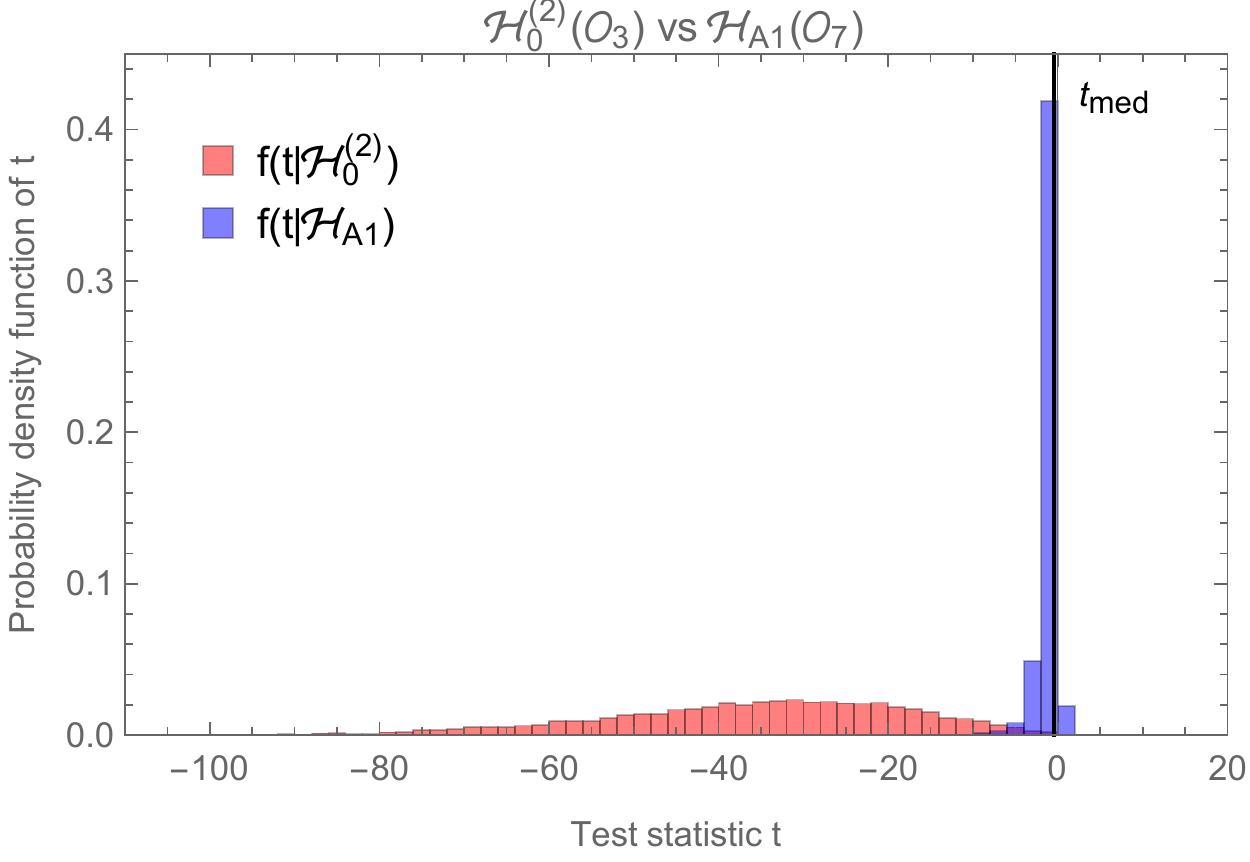}
\caption{Histograms representing the probability density function of the log-likelihood ratio $t$ sampled under the null hypothesis, $\mathcal{O}_1$ Tyranny in the left panel ($\mathcal{H}_{0}^{(1)}$, pink) and $\mathcal{O}_3$ Tyranny in the right panel ($\mathcal{H}_{0}^{(2)}$, pink), as well as under the alternative hypothesis, $\mathcal{H}_{A1}$ (P-violations, blue) both in the left and right panels.~In both panels, the probability density function of $t$ under the alternative hypothesis is rather narrow and peaks at a value of $t$ just below zero.~However, the probability density function of $t$ peaks at large negative values when data are sampled from the null hypothesis (both in the left and right panel).~This evident separation of histograms implies that the functions $f(t|\mathcal{H}_0^{(1)})$ ($f(t|\mathcal{H}_0^{(2)})$) and $f(t|\mathcal{H}_{A1})$ are clearly distinguishable if about 10 (180) signal events are recorded in a xenon detector.}
\label{fig:histograms}
\end{figure}

We now move from the qualitative considerations to the calculation of the number of DM signal events required to reject P and CP invariance in favour of P, or P- and CP-violations.~This calculation is based on the Monte Carlo simulation of probability density functions for the log-likelihood test statistic under the hypotheses of interest.~As illustrative examples, Fig.~\ref{fig:histograms} reports two selected Monte Carlo generated probability density functions for $t$.~The  left panel shows the probability density functions $f(t|\mathcal{H}_0)$ for $\mathcal{H}_0$ = $\mathcal{H}_0^{(1)}$ ($\mathcal{O}_1$ Tyranny) and $f(t|\mathcal{H}_{A1})$.~We obtain both histograms from about $2\times10^4$ Monte Carlo simulations of nuclear recoil events setting the DM particle mass to the reference value of 30 GeV.~Under both hypotheses, we set the free coupling constants and the exposure to a value producing about 10 signal events in a xenon detector.~In this particular case, the statistical significance with which the $\mathcal{O}_1$ Tyranny can be rejected in favour of P-violations in spin-0 DM-nucleus interactions is of about 3.1.~Similarly, the right panel of Fig.~\ref{fig:histograms} shows the probability density functions $f(t|\mathcal{H}_0)$ for $\mathcal{H}_0$ = $\mathcal{H}_0^{(2)}$ ($\mathcal{O}_3$ Tyranny) and $f(t|\mathcal{H}_{A1})$.~In this second example, we set the free coupling constants and the experimental exposure to values producing about 180 signal events assuming a xenon detector, and the associated statistical significance is again $\mathcal{Z}\simeq3.1$.~As one can see from Fig.~\ref{fig:histograms} (left and right panel), the probability density function of $t$ under the alternative hypothesis $\mathcal{H}_{A1}$ (the same would apply to $\mathcal{H}_{A2}$) is rather narrow and peaks at a value of $t$ just below zero.~This means that when data are sampled under $\mathcal{H}_{A1}$ the likelihood ratio in Eq.~(\ref{eq:t}) is about one in most of the Monte Carlo simulations, with a slight preference for values just below one.~Notice that $t=0$ implies that null and alternative hypotheses fit the data simulated under the alternative hypothesis equally well, whereas $t<0$ indicates that the null hypothesis (the wrong model in this case) can fit the data sampled from $\mathcal{H}_{A1}$ even better than the alternative hypothesis itself.~This behaviour is expected and reflects the fact that the null hypothesis has two free parameters, $c_1^N$ and $c_3^N$, and is therefore more flexible than $\mathcal{H}_{A1}$, where the only free parameter is $c_7^N$.~At the same time, the left and right panels of Fig.~\ref{fig:histograms} show that the probability density function of $t$ peaks at large negative values when data are sampled from the null hypothesis (we find a peak at large negative values of $t$ independently of the specific realisation of $\mathcal{H}_0$).~This implies that $f(t|\mathcal{H}_0^{(1)})$ ($f(t|\mathcal{H}_0^{(2)})$) and $f(t|\mathcal{H}_{A1})$ are clearly distinguishable if about 10 (180) signal events are recorded in a xenon detector, as in the present examples.~We find similar probability density functions for all pairs of hypotheses in Tab.~\ref{tab:h}.

\begin{figure}
\centering
\includegraphics[width=0.95\textwidth]{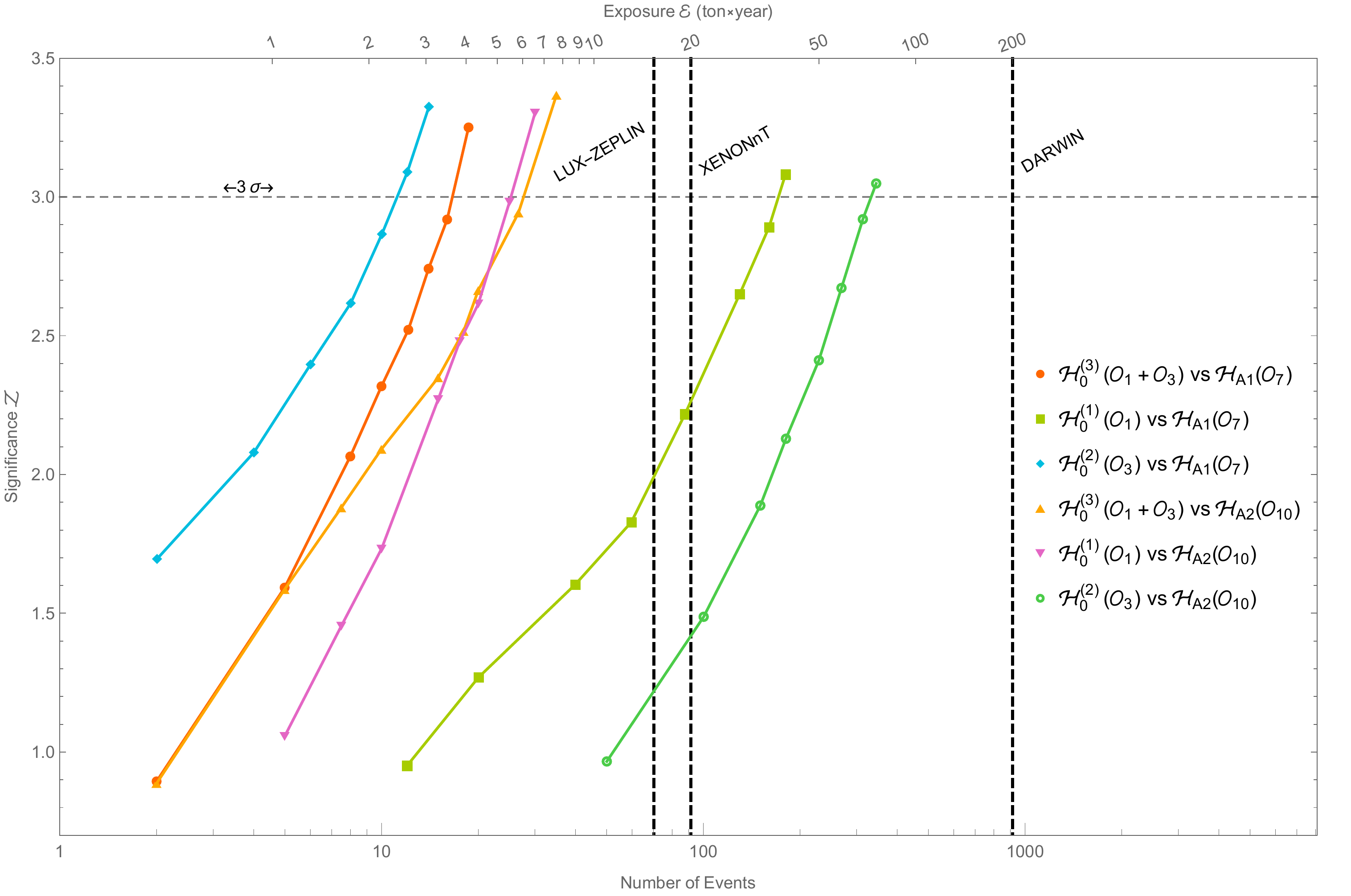}
\caption{Statistical significance for rejecting P and CP in spin-0 DM-nucleus interactions as a function of the observed number of DM signal events in a xenon detector.~We statistically compare three P and CP preserving hypotheses, $\mathcal{H}_0^{(1)}$, $\mathcal{H}_0^{(2)}$ and $\mathcal{H}_0^{(3)}$, against two alternative hypotheses, $\mathcal{H}_{A1}$ (P-violations) and $\mathcal{H}_{A2}$ (P- and CP-violations).~We set the DM particle mass to the reference value of $30$~GeV.}
\label{fig:all}
\end{figure}

We now move to one of the main results of this work.~Fig.~\ref{fig:all} shows a plot of the statistical significance for rejecting P and CP preserving DM-nucleus interactions under $\mathcal{O}_1$ Tyranny ($\mathcal{H}_0^{(1)}$), $\mathcal{O}_3$ Tyranny ($\mathcal{H}_0^{(2)}$), and Democracy ($\mathcal{H}_0^{(3)}$) in favour of P-violations ($\mathcal{H}_{A1}$) or P- and CP-violations ($\mathcal{H}_{A1}$) in spin-0 DM-nucleus interactions as a function of the number of DM signal events observed in a xenon detector.~In Fig.~\ref{fig:all}, different lines correspond to distinct pairs of tested hypotheses (see Tab.~\ref{tab:h}) and $m_\chi=30$~GeV.~As expected based on the predicted nuclear recoil energy spectra (see Fig.~\ref{fig:spectra}), rejecting the $\mathcal{O}_3$ Tyranny in favour of P violating DM-nucleus interactions is relatively simple, requiring only about 10 DM signal events to reach a statistical significance corresponding to three standard deviations.~On the contrary, rejecting the $\mathcal{O}_3$ Tyranny with a statistical significance of $\mathcal{Z}=3$ in favour of P- and CP-violations in DM-nucleus interactions is rather challenging and requires more than 300 DM signal events.

The top x-axis in Fig.~\ref{fig:all} shows the exposure corresponding to the number of signal events in the bottom x-axis when coupling constants are set to match the current XENON1T 90\% C.L~exclusion limit on $c_1^N$~\cite{aprile:2018dbl}.~Dashed vertical lines indicate the expected exposure for LZ~\cite{Akerib:2018lyp}, XENONnT~\cite{Aprile:2015uzo} and DARWIN~\cite{Aalbers:2016jon}.~This figure demonstrates that rejecting P and CP invariance in spin-0 DM-nucleus interactions is within the reach of next generation DM direct detection experiments.

\begin{figure}
\centering
\includegraphics[width=0.49\textwidth]{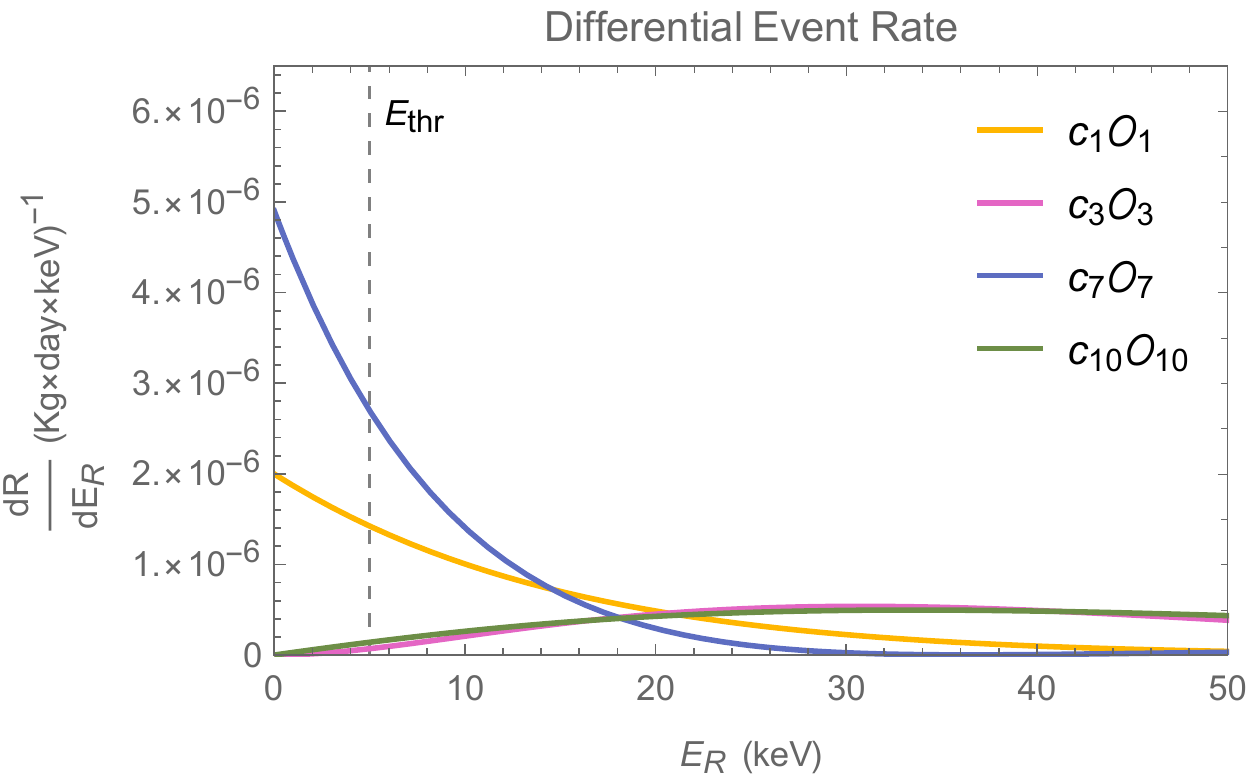}
\includegraphics[width=0.5\textwidth]{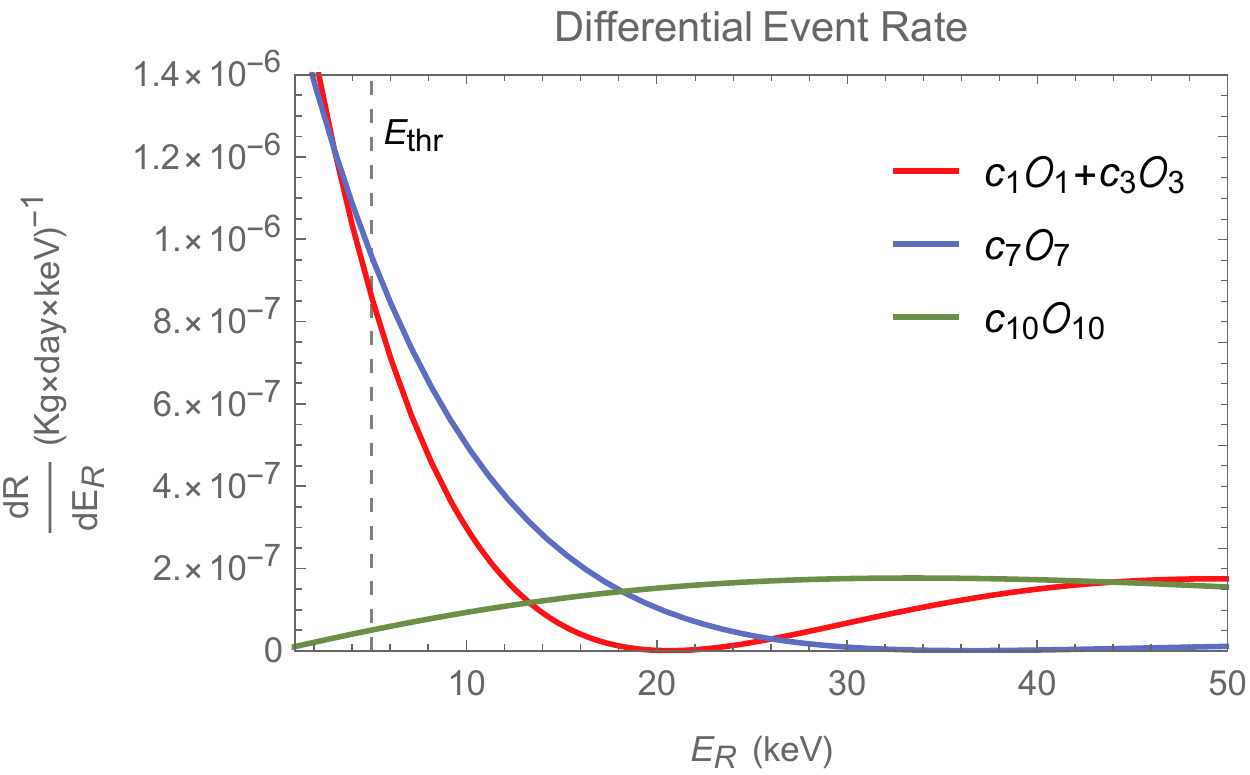}
\includegraphics[width=0.49\textwidth]{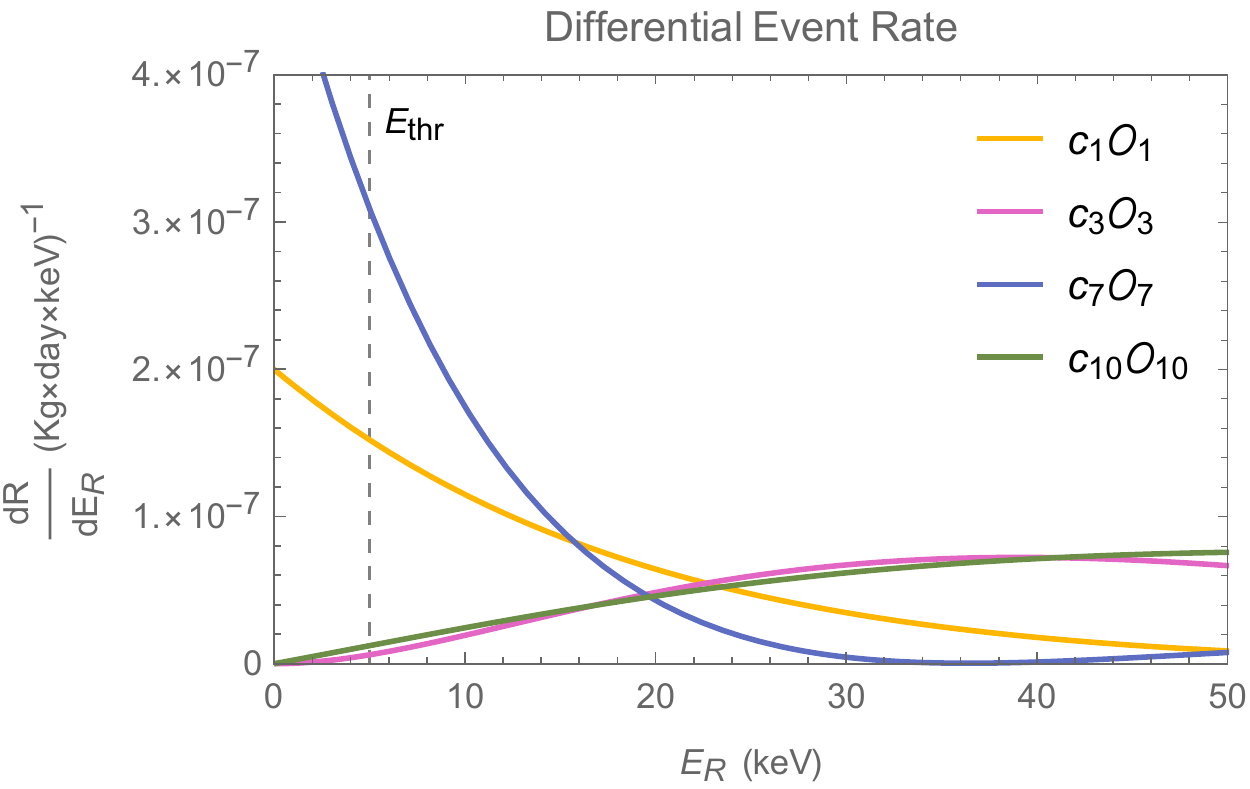}
\includegraphics[width=0.49\textwidth]{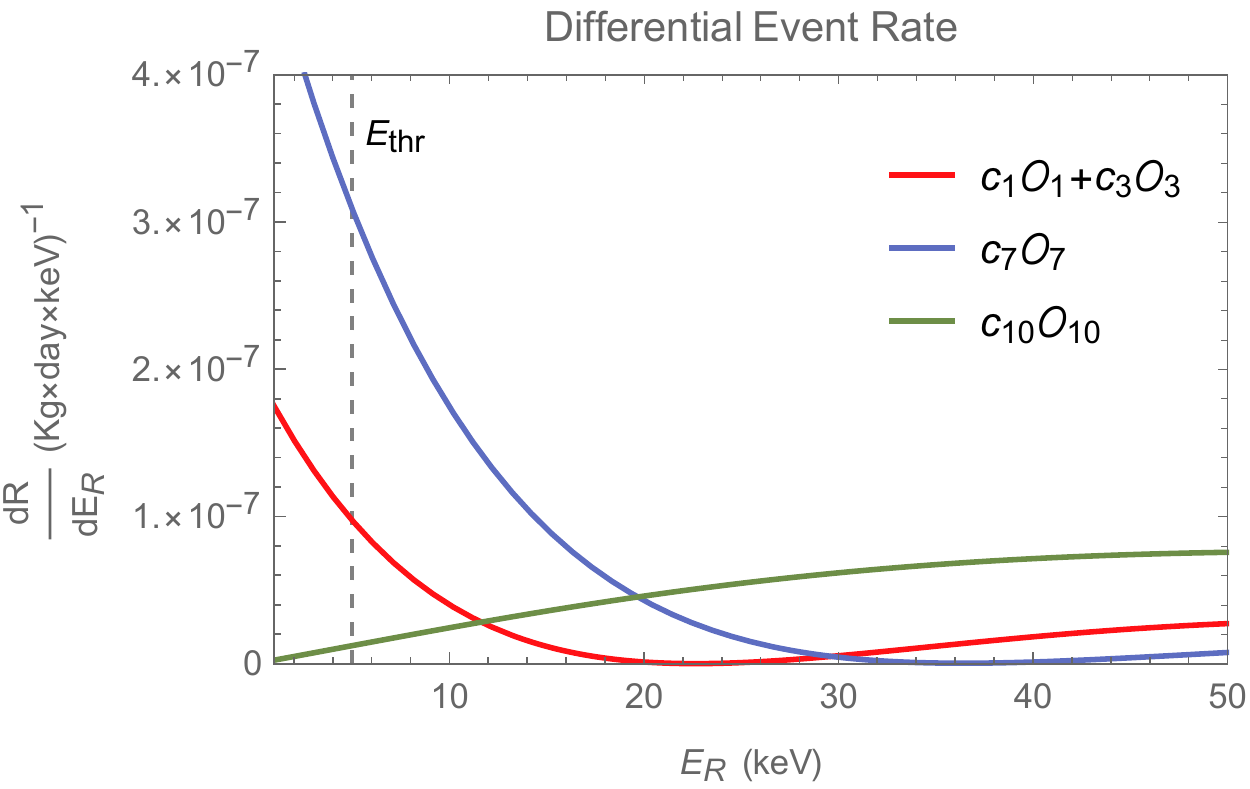}
\caption{Nuclear recoil energy spectra for different DM particle masses.~{\it Top left panel}.~Same as Fig.~\ref{fig:spectra}, but now for $m_\chi=100$~GeV.~{\it Top right panel}.~Same as Fig.~\ref{fig:spectra2}, now with $m_\chi=100$~GeV.~{\it Bottom left panel}.~Same as Fig.~\ref{fig:spectra}, now for $m_\chi=1$~TeV.~{\it Bottom right panel}.~Same as Fig.~\ref{fig:spectra2}, but now with $m_\chi=1$~TeV.
\label{fig:spectra3}}
\end{figure}

So far, we focused on 30 GeV as a benchmark value for the DM particle mass, because the strongest exclusion limits on the coupling constants for DM-nucleon interactions set by the XENON1T experiment are associated with comparable values of $m_\chi$~\cite{aprile:2018dbl}.~We now extend our analysis to other DM particle masses.~For $m_\chi$ below 30~GeV, nuclear recoil energy spectra corresponding to different operators are similar above the assumed energy threshold (5 keV).~Consequently, below $m_\chi=30$~GeV rejecting P and CP invariance is extremely difficult based on the analysis of nuclear recoil data.~On the contrary, when increasing the DM particle mass, differences in the nuclear recoil energy spectra persist, as one can see from Fig.~\ref{fig:spectra3} where we focus on two representative cases corresponding to $m_\chi=100$~GeV (top panels) and and $m_\chi=1$~TeV (bottom panels), respectively.~Quantitatively, we find that P and CP invariance can be rejected in favour of P, or P- and CP-violations with a comparable number of signal events for $m_\chi$ above our benchmark value of 30~GeV (see Tab.~\ref{tab:mass}).~This means that over a wide range of DM masses, it is in fact feasible to test for P- or CP-violation in future direct detection experiments.

In this work we assumed that the DM particle has spin-0, but it should be clear that an analogous analysis can be done for spin-1/2 DM (or spin-1 DM). Such a study would, however, be significantly more involved and time-consuming than the one presented here, due to the larger number of operators that can contribute to the interaction.  In fact, both the number of operators that violate P and CP as well the number of operators that are invariant under them increase in such a case.  We will address this possibility in a separate work.

\section{Conclusions}
\label{sec:conclusions}

We computed the number of DM signal events required to reject P and CP invariance in favour of P, or P- and CP-violations in the scattering of spin-0 DM particles by nuclei in a xenon detector.~We considered three distinct realisations of the P and CP preserving hypothesis: $\mathcal{O}_1$ Tyranny ($\mathcal{H}_0^{(1)}$), $\mathcal{O}_3$ Tyranny ($\mathcal{H}_0^{(2)}$), and Democracy ($\mathcal{H}_0^{(3)}$).~We denoted the P violating hypothesis by $\mathcal{H}_{A1}$ and the P and CP violating scenario by $\mathcal{H}_{A2}$.~See Tab.~\ref{tab:h} for a summary of the hypotheses we compared.~We performed this calculation by using the log-likelihood ratio as a test statistic and Monte Carlo simulations of nuclear recoil events to compute the median $p$-value for rejecting $\mathcal{H}_{0}^{(i)}$, $i=1,2,3$.~The outcome of this analysis is summarised in Fig.~\ref{fig:all}.

We found that the number of DM signal events required to reject P and CP invariance depends on how we model the hypothesis of P and CP preserving spin-0 DM-nucleus interactions, and on whether the alternative hypothesis implies P, or both P- and CP-violations.~For example, rejecting P and CP invariance under the ``$\mathcal{O}_3$ Tyranny'' in favour of P violating DM-nucleus interactions is relatively simple, as it requires only about 10 DM signal events to reach a statistical significance of $\mathcal{Z}=3$.~In contrast, rejecting the $\mathcal{O}_3$ Tyranny with a statistical significance corresponding to three standard deviations in favour of P- {\it and} CP-violations in DM-nucleus interactions is rather challenging, as it requires more than 300 DM signal events.~The other hypotheses we compared in this analysis require a number of DM signal events to reach the threshold $\mathcal{Z}=3$ that lies in between about 10 and 300 (see Fig.~\ref{fig:all} for a summary).~Expressing the number of signal events associated with a given significance in terms of a corresponding experimental exposure, we found that rejecting P and CP invariance in spin-0 DM-nucleus interactions is within reach of next generation DM direct detection experiments.~Qualitatively, our results remain valid for DM particle masses above about 30 GeV.

\begin{table}
\renewcommand{\arraystretch}{1.8}
\centering
\begin{tabular}{lcccc}
\hline
DM mass & Null hypothesis & Alternative & Significance & Number of events  \\
\hline
\hline
$m_\chi=30$~GeV & $\mathcal{H}_{0}^{(1)}$ & $\mathcal{H}_{A1}$ & $\mathcal{Z}=3.08$ &  180.2  \\
$m_\chi=30$~GeV & $\mathcal{H}_{0}^{(1)}$ & $\mathcal{H}_{A2}$ & $\mathcal{Z}=2.98$ & 25.0  \\
$m_\chi=100$~GeV & $\mathcal{H}_{0}^{(1)}$ & $\mathcal{H}_{A1}$ & $\mathcal{Z}=3.00$ & 227.8  \\
$m_\chi=100$~GeV & $\mathcal{H}_{0}^{(1)}$ & $\mathcal{H}_{A2}$ & $\mathcal{Z}=3.18$ & 99.6  \\
$m_\chi=1$~TeV & $\mathcal{H}_{0}^{(1)}$ & $\mathcal{H}_{A1}$ & $\mathcal{Z}=3.11$ & 48.9 \\
$m_\chi=1$~TeV & $\mathcal{H}_{0}^{(1)}$ & $\mathcal{H}_{A2}$ & $\mathcal{Z}=3.28$ & 20.4  \\
\hline
\end{tabular}
\caption{Statistical significance (fourth column) and corresponding number of DM signal events (fifth column) to reject the null hypothesis in the second column in favour of the alternative hypothesis in the third column for the DM particle mass in the first column.~This table partly extends Fig.~\ref{fig:all} to other masses.}
\label{tab:mass}
\end{table}
Summarising, our analysis shows that a signal in a direct detection experiment can be used to extract information of the discrete symmetries that underly the spin-0 DM-nucleus scattering, and in this way provide decisive clues about the fundamental nature of the DM particle. 

\acknowledgments 
It is a pleasure to thank Timon Emken for the many valuable discussions and for his help with the numerical implementation of the Monte Carlo simulations performed in this work.~RC acknowledges support from an individual research grant from the Swedish Research Council, dnr. 2018-05029.~The research presented here made use of the computer programme Wolfram Mathematica~\cite{Mathematica}.

\appendix
\section{T-matrix element derivation}
\label{sec:T}
Let us consider the scattering of a DM particle by an atomic nucleus consisting of $A$ bound nucleons.~We denote the initial state for this process by the tensor product, $|i\rangle= |T_i\rangle \otimes a^{\dagger}_{\mathbf {p}}|0\rangle$, between a nuclear state $|T_i\rangle$ and a DM particle state, $a^{\dagger}_{\mathbf {p}}|0\rangle$, where $a^{\dagger}_{\mathbf {p}}$ ($a_{\mathbf {p}}$) is the creation (annihilation) operator of spin-0 DM quanta with three-dimensional momentum $\mathbf{p}$ and $|0\rangle$ is the vacuum.~Similarly, we denote the final state for this process by $|f\rangle= |T_f\rangle \otimes a^{\dagger}_{\mathbf {p}'}|0\rangle$, where $|T_f\rangle$ is the state of the outgoing nucleus and $\mathbf{p}'$ the three-dimensional momentum of the final state DM particle.~Notice that the norm of the DM particle state $a^{\dagger}_{\mathbf {p}}|0\rangle$ is divergent and equal to $V\equiv (2\pi)^3 \delta^{(3)}(0) = \langle 0 |a_{\mathbf {p}} a^{\dagger}_{\mathbf {p}}|0\rangle$.~However, any measurable quantity will be independent of the volume $V$.~The S-matrix element associated with this scattering process can be written as follows
\begin{align}
S_{fi} &= - \int {\rm d}^3 \mathbf{x}_1 \int {\rm d}^3 \mathbf{x}_2 \int_{-\infty}^{+\infty} {\rm d} t_1 \int_{-\infty}^{t_1} {\rm d} t_2 \, \langle f | \mathscr{H}(t_1,\mathbf{x}_1)  \mathscr{H}(t_2,\mathbf{x}_2)  |i \rangle \,,
\label{eq:S1}
\end{align}
where $\mathscr{H}(t_1,\mathbf{x}_1)$ ($\mathscr{H}(t_2,\mathbf{x}_2)$) is the Hamiltonian density for DM-mediator and $A$ nucleons-mediator interactions at the space-time point $x_1=(t_1,\mathbf{x}_1)$ ($x_2=(t_2,\mathbf{x}_2)$).~For example, in the case of spin-0 DM-nucleon interactions mediated by a scalar particle, $\mathscr{H}$ would read as follows
\begin{align}
\mathscr{H} = g_1 m_\chi\phi  \chi^\dagger \chi + h_1 \phi  \sum_{i=1}^A \bar{\psi}_{N_i} \psi_{N_i} \,,
\end{align}
where $g_1$ and $h_1$ are coupling constants, $m_\chi$ is the DM particle mass, $\chi$ ($\phi$) is the DM (mediator) scalar field and, finally, $\psi_{N_i}$ is a spinor field for the $i$-th nucleon.~By inserting the identity operator $1=\sum_m |m \rangle \langle m|$ between the two Hamiltonian densities in (\ref{eq:S1}), where $|m\rangle$ are eigenstates of the Hamiltonian, $H_0$, for the $A$ nucleons-DM system with $\mathscr{H}$ set to zero, we obtain
\begin{align}
S_{fi} &=  - \sum_{m} \int {\rm d}^3 \mathbf{x}_1 \int {\rm d}^3 \mathbf{x}_2 \int_{-\infty}^{+\infty} {\rm d} t_1 \int_{-\infty}^{t_1} {\rm d} t_2 \, \langle f | \mathscr{H}(t_1,\mathbf{x}_1) |m \rangle \langle m| \mathscr{H}(t_2,\mathbf{x}_2)  |i \rangle \,.
\label{eq:S2}
\end{align}
We now translate $\mathscr{H}(t_1,\mathbf{x}_1)$ and $\mathscr{H}(t_2,\mathbf{x}_2)$ to $t_1=t_2=0$ via the time translation operators $e^{-i H_0 t_1}$ and $e^{-i H_0 t_2}$, respectively.~Denoting by $E_i$, $E_f$ and $E_m$ the solutions to the eigenvalue equations $H_0 | i \rangle = E_i | i \rangle$, $H_0 | f \rangle = E_f | f \rangle$ and $H_0 | m \rangle = E_m | m \rangle$, we can rewrite Eq.~(\ref{eq:S2}) as follows
\begin{align}
S_{fi} &= - \sum_{m} \int {\rm d}^3 \mathbf{x}_1 \int {\rm d}^3 \mathbf{x}_2  \, \langle f | \mathscr{H}(0,\mathbf{x}_1) |m \rangle \langle m| \mathscr{H}(0,\mathbf{x}_2)  |i \rangle \int_{-\infty}^{+\infty} {\rm d} t_1 \,e^{i (E_f-E_m) t_1} \nonumber\\
&\times \int_{-\infty}^{t_1} {\rm d} t_2 \, e^{i (E_m-E_i-i\epsilon) t_2} \,,
\end{align}
where we introduce the parameter $\epsilon$ (to be set to zero at the end of the calculation) to ensure the convergence of the second time integral in the $t_2\rightarrow -\infty$ limit.~Performing the time integrals, we find
\begin{align}
S_{fi} &=  - 2\pi i \delta(E_f - E_i) \int {\rm d}^3 \mathbf{x}_1 \int {\rm d}^3 \mathbf{x}_2  \, \langle f | \mathscr{H}(0,\mathbf{x}_1)  \frac{1}{E_i-H_0+i\epsilon} \mathscr{H}(0,\mathbf{x}_2)  |i \rangle \,.
\label{eq:S3}
\end{align}
From Eq.~(\ref{eq:S3}) one can read the explicit form of the T-matrix element associated with $S_{fi}$ by noticing that
\begin{align}
S_{fi}\equiv -2\pi \delta(E_f-E_i)\,iT_{fi}\,.
\label{eq:ST}
\end{align}
By inserting in Eq.~(\ref{eq:S3}) the identity operator expressed in terms of free initial 
A-nucleon states,
\begin{align}
&\sum_{s_1,\dots,s_A}\left( \prod_{i=1}^{A} \int \frac{{\rm d}^3 \mathbf{k}_i}{(2 \pi)^3} \right) | \mathbf{k}_1,s_1;\dots, \mathbf{k}_A, s_A\rangle \langle  \mathbf{k}_1,s_1;\dots, \mathbf{k}_A ,s_A|= 1 \,, 
\end{align}
and final A-nucleon states
\begin{align}
&\sum_{s'_1,\dots,s'_A}\left( \prod_{i=1}^{A} \int \frac{{\rm d}^3 \mathbf{k}'_i}{(2 \pi)^3} \right) | \mathbf{k}'_1,s'_1;\dots, \mathbf{k}'_A, s'_A\rangle \langle  \mathbf{k}'_1,s'_1;\dots, \mathbf{k}'_A ,s'_A|= 1 \,, 
\end{align}
where $| \mathbf{k}_i,s_i \rangle = | \mathbf{k}_i\rangle \times |s_i \rangle$, $\mathbf{k}_i$ and $s_i$  ($\mathbf{k}'_i$ and $s_i'$), $i=1,\dots,A$ are the initial (final) constituent nucleon three-dimensional momenta and spin third component, and $\sum_{s_i} |s_i\rangle \langle s_i|=1$, we obtain
\begin{align}
S_{fi} &=  - 2\pi i \delta(E_f - E_i)  \sum_{s_1',\dots,s_A'}\sum_{s_1,\dots,s_A}
\left( \prod_{i=1}^{A} \int \frac{{\rm d}^3 \mathbf{k}_i}{(2 \pi)^3} \right) \left( \prod_{i=1}^{A} \int \frac{{\rm d}^3 \mathbf{k}'_i}{(2 \pi)^3} \right) \, V^A\,T^{\rm free}_{fi}\nonumber\\
&\times \psi^*_f(\mathbf{k}'_1,s'_1;\dots; \mathbf{k}'_A,s'_A) \psi_{in}(\mathbf{k}_1,s_1;\dots; \mathbf{k}_A,s_A) \,.
\label{eq:S4}
\end{align}
Here, we denote by $\psi_{in}(\mathbf{k} _1,s_1;\dots; \mathbf{k}_A,s_A)=\langle   \mathbf{k} _1,s_1;\dots; \mathbf{k}_A,s_A |T_i \rangle$ the initial nuclear wave function and by $\psi^*_f(\mathbf{k}'_1,s'_1;\dots; \mathbf{k}'_A,s'_A)=\langle T_f | \mathbf{k}'_1,s_1';\dots; \mathbf{k}'_A,s_A' \rangle$, the associated final state nuclear wave function.~Furthermore, we denote by $T^{\rm free}_{fi}$ the T-matrix element for the scattering of a DM particle by a nucleon in a system of $A$ free nucleons.~By analogy with $(\ref{eq:S3})$, we can write
\begin{align}
T^{\rm free}_{fi} &= V^{-A} \int {\rm d}^3 \mathbf{x}_1 \int {\rm d}^3 \mathbf{x}_2 \,  \langle  \mathbf{k}' _1,s_1;\dots, \mathbf{k}'_A,s_A;\mathbf{p}' |  \mathscr{H}(0,\mathbf{x}_1)   \frac{ 1 }{\tilde{E}_i-\tilde{H}_0+i\epsilon} \nonumber\\
&\times \mathscr{H}(0,\mathbf{x}_2)   | \mathbf{k}_1,s_1\dots, \mathbf{k}_A, s_A;\mathbf{p}\rangle \,,
\label{eq:T}
\end{align}
where, $| \mathbf{k}_1,s_1;,\dots, \mathbf{k}_A,s_A; \mathbf{p}\rangle=| \mathbf{k}_1,\dots, \mathbf{k}_A \rangle \times  | s_1,\dots, s_A \rangle \times a^\dagger_\mathbf{p} |0\rangle$  and the A-nucleon state $| \mathbf{k}_1,\dots, \mathbf{k}_A \rangle$ is normalised as follows $\langle \mathbf{k}_1,\dots, \mathbf{k}_A | \mathbf{k}_1,\dots, \mathbf{k}_A \rangle = V^A$. Here, $\tilde{H}_0$ is the Hamiltonian of the $A$ free nucleon-DM system with $\mathscr{H}$ set to zero and $\tilde{E}_i$ is the solution to $\tilde{H}_0  | \mathbf{k}_1,\dots, \mathbf{k}_A ,\mathbf{p}\rangle  = \tilde{E}_i  | \mathbf{k}_1,\dots, \mathbf{k}_A ,\mathbf{p}\rangle $.~Notice that, if $| \tilde{m} \rangle$ is an eigenstate of $\tilde{H}_0$ and $| m \rangle$ is an eigenstate of $H_0$ such that $\tilde{E}_m-E_m=E_{b}$, where $E_{b}$ is the nucleon binding energy, then $\langle \tilde{m} |(\tilde{E}_i-\tilde{H}_0+i\epsilon)^{-1}| \tilde{m} \rangle = \langle m |(E_i-H_0+i\epsilon)^{-1}| m \rangle=(E_i-E_m+i\epsilon)^{-1}$, as the binding energy contribution to the energies $E_i$ and $E_m$ cancels in the difference $E_i-E_m$.~In order to investigate the properties of the S-matrix element $S_{fi}$ under P and CP, it is convenient to relate the T-matrix element $T^{\rm free}_{fi}$ in Eq.~(\ref{eq:T}) to the amplitude $M_{\chi N_j}$ for DM scattering by the $j$-th nucleon in the sample of $A$ free nucleons of initial (final) momenta $\mathbf{k}_1,\dots,\mathbf{k}_A$ ($\mathbf{k}'_1,\dots,\mathbf{k}'_A$).~We obtain this relation by matching the S-matrix element, $S^{\rm free}_{fi}$, for the $A+1 \rightarrow A+1$ process where DM scatters on one free nucleon while leaving the remaining $A-1$ free nucleons unscattered, 
\begin{align}
S^{\rm free}_{fi}=(2\pi)^4 \delta(\tilde{E}_f-\tilde{E}_i) \sum_{j=1}^{A} \delta^{(3)}(\mathbf{p}'+\mathbf{k}'_j-\mathbf{p}-\mathbf{k}_j) \frac{V}{\sqrt{16 E_{\mathbf{p}} E_{\mathbf{k}_j} E_{\mathbf{p}'} E_{\mathbf{k}'_j} V^4}}\,i M^{(j)}_{\chi N_1,\dots,N_A}\,, 
\label{eq:Sfree}
\end{align}
on to the definition $S^{\rm free}_{fi}\equiv- (2\pi) \delta(\tilde{E}_f-\tilde{E}_i)\,iT^{\rm free}_{fi}$.~Here, $M^{(j)}_{\chi N_1,\dots,N_A}$ is the scattering amplitude associated with $S^{\rm free}_{fi}$, for which an explicit expression as a function of $M_{\chi N_j}$ reads as follows,
\begin{align}
M^{(j)}_{\chi N_1,\dots N_A} = M_{\chi N_j}\prod_{i=1 \atop i\ne j}^{A}
V^{-1}(2\pi)^3\delta^{(3)} (\mathbf{k}'_{i}-\mathbf{k}_{i}) \delta^{s'_i s_i}\,,
\end{align}
where $V^{-1}\delta^{(3)} (\mathbf{k}'_{i}-\mathbf{k}_{i})$, $M_{\chi N_j}$ (see Eq.~(\ref{eq:M})) and therefore $M^{(j)}_{\chi N_1,\dots,N_A}$ are dimensionless.~The $V$ factor in the numerator of Eq.~(\ref{eq:Sfree}) arises when rewriting $a^\dagger_\mathbf{p}|0\rangle$ as $\sqrt{V}a^\dagger_\mathbf{p}|0\rangle/\sqrt{V}$.~Indeed, the scattering amplitudes $M_{\chi N_j}$ and $M^{(j)}_{\chi N_1,\dots,N_A}$ are dimensionless only if evaluated between unit-normalised DM particle states of the type $a^\dagger_\mathbf{p}|0\rangle/\sqrt{V}$, and analogously for nucleons.~This leads us to
\begin{align}
T^{\rm free}_{fi} = -\sum_{j=1}^{A}(2\pi)^3 \delta^{(3)}(\mathbf{p}'+\mathbf{k}_j' - \mathbf{p}-\mathbf{k}_j) \frac{V M_{\chi N_j}}{\sqrt{16 E_{\mathbf{p}} E_{\mathbf{k}_j} E_{\mathbf{p}'} E_{\mathbf{k}'_j} V^4}} \, \prod_{i=1 \atop i\ne j}^{A}
V^{-1}(2\pi)^3\delta^{(3)} (\mathbf{k}'_{i}-\mathbf{k}_{i})\delta^{s'_i s_i}
\label{eq:Tfree}
\end{align}
Combining Eqs.~(\ref{eq:Tfree}) and (\ref{eq:S4}) with Eq.~(\ref{eq:ST}), we obtain our final expression for the T-matrix element $T_{fi}$,
\begin{align}
T_{fi} = -\sum_{j=1}^{A} \left( \prod_{i=1}^{A} \int \frac{{\rm d}^3 \mathbf{k}_i}{(2 \pi)^3} \right) \frac{\psi^*_f(\mathbf{k}_1^j,
\dots,\mathbf{k}^j_A) \psi_{in}(\mathbf{k}_1,\dots, \mathbf{k}_A)}{\sqrt{16 E_{\mathbf{p}} E_{\mathbf{k}_j} E_{\mathbf{p}'} E_{\mathbf{k}'_j}}}\,M_{\chi N_j}\,,
\label{eq:Tfinal}
\end{align}
where $\mathbf{k}_m^j=\mathbf{k}_m+\mathbf{q}$ for $m=j$ and $\mathbf{k}_m^j=\mathbf{k}_m$ otherwise.~In order to simplify the notation, in Eq.~(\ref{eq:Tfinal}) we omit the nucleon spin indices in the scattering amplitude $\mathcal{M}_{\chi N_j}$ and in the initial and final nuclear wave functions.~Furthermore, the two sums $\sum_{s_1',\dots,s_A'}$ and $\sum_{s_1,\dots,s_A}$ are understood.

\providecommand{\href}[2]{#2}\begingroup\raggedright\endgroup

\end{document}